%% file: main.tex
\documentclass{article}
\usepackage{PRIMEarxiv}
\usepackage[utf8]{inputenc} 
\usepackage[T1]{fontenc}    
\usepackage{hyperref}       
\usepackage{url}            
\usepackage{booktabs}       
\usepackage{amsfonts}       
\usepackage{nicefrac}       
\usepackage{microtype}      
\usepackage{lipsum}
\usepackage{fancyhdr}       
\usepackage{graphicx}       
\graphicspath{{media/}}     
\usepackage{times}  
\usepackage{helvet}  
\usepackage{courier}  
\usepackage{natbib}  
\usepackage{caption} 
\usepackage{algorithm}
\usepackage{graphicx}
\usepackage{textcomp}
\usepackage{caption}
\usepackage{subcaption}
\usepackage{booktabs}
\usepackage{wrapfig}
\usepackage{amsthm}
\usepackage[switch]{lineno}
\usepackage{multirow}
\usepackage{bbm}
\usepackage{booktabs}
\usepackage[noend]{algpseudocode}
\usepackage{mathtools}
\usepackage{mathrsfs}
\usepackage{comment}
\usepackage{outlines}
\usepackage{amsmath,amsfonts}

\makeatletter
\def\@captype{table}
\makeatother

\title{Does AI-Assisted Fact-Checking Disproportionately Benefit Majority Groups Online?}

\begin{document}

\author{
  Terrence Neumann \thanks{Equal contribution}\\
  University of Texas at Austin \\
  \texttt{Terrence.Neumann@mccombs.utexas.edu} \\
   \And
  Nicholas Wolczynski $^*$\\
   University of Texas at Austin \\
   \texttt{nicholas@mccombs.utexas.edu} \\
}

\maketitle

\input{abstract}

\section{Introduction}
\input{introduction.tex}

\section{Related Work}
\input{related_work.tex}

\section{Approach}
\label{sec:approach}
\input{approach_edits.tex}

\section{Experiments}
\label{sec:experiment}
\input{experiment.tex}

\section{Discussion \& Conclusions}
\input{conclusion.tex}

\bibliographystyle{main.bst}
\bibliography{main.bib}

\clearpage

\appendix
\input{appendix.tex}

\end{document}

%% file: abstract.tex
\begin{abstract}
In recent years, algorithms have been incorporated into fact-checking pipelines. They are used not only to flag previously fact-checked misinformation, but also to provide suggestions about which trending claims should be prioritized for fact-checking - a paradigm called `check-worthiness.' While several studies have examined the accuracy of these algorithms, none have investigated how the benefits from these algorithms (via reduction in exposure to misinformation) are distributed amongst various online communities. In this paper, we investigate how diverse representation across multiple stages of the AI development pipeline affects the distribution of benefits from AI-assisted fact-checking for different online communities. We simulate information propagation through the network using our novel Topic-Aware, Community-Impacted Twitter (TACIT) simulator on a large Twitter followers network, tuned to produce realistic cascades of true and false information across multiple topics. Finally, using simulated data as a test bed, we implement numerous algorithmic fact-checking interventions that explicitly account for notions of diversity. We find that both representative \textit{and} egalitarian methods for sampling and labeling check-worthiness model training data can lead to network-wide benefit concentrated in majority communities, while incorporating diversity into how fact-checkers use algorithmic recommendations can actively reduce inequalities in benefits between majority and minority communities. These findings contribute to an important conversation around the responsible implementation of AI-assisted fact-checking by social media platforms and fact-checking organizations.
\end{abstract}

%% file: introduction.tex
Misinformation continues to propagate widely on social media, causing significant uncertainty, disagreement, and, at times, violence around critical events. Recent efforts to combat the spread of falsehoods online have involved human-AI collaboration in which AI is trained by crowd-workers to determine which content should be further evaluated by professional fact-checkers \citep{allen2022birdwatch, SaeedCrowdsource22}. After claims are fact-checked, natural language processing (NLP) is used to spot repeated mentions of misinformation and reduce their circulation online \citep{meta_for_media_2021, fullfact}. This approach has enabled scalable misinformation detection online, with Facebook labeling hundreds of millions of posts related to national elections and public health as potentially misleading \citep{lerman_covid_2020, lerman_kelly_2020}.

However, there are growing concerns about mis- and disinformation ``slipping through the cracks'' and disproportionately affecting minority communities online, with a Senate Intelligence report citing that ``no single group of Americans was targeted ... more than African Americans. By far, race and related issues were the preferred target of the information warfare campaign designed to divide the country in 2016'' \cite[pg.6]{congress2020}. Similarly, we see concerns in Europe regarding the continued circulation of persistent false narratives of particular migrant communities \citep{szakacs2021impact}.

To date, the majority of work on fact-checking algorithms used to identify ``check-worthy'' claims circulating online has focused on designing models that have high accuracy on test sets of previously fact-checked claims \citep{checkthat, twitter_metadata, claimbuster, automated_claim_detection}. However, prior work has not considered whether these AI tools provide equal benefit to all people when deployed. We ask whether algorithms used to assist fact-checkers tend to identify falsehoods that impact majority groups while allowing misinformation that impacts minority groups to proliferate relatively unimpeded. Additionally, we investigate whether explicitly incorporating notions of diversity (as discussed in \citet{fazelpour2022diversity}) into the training data and AI-assisted fact-checking workflow improves overall outcomes and affects the distribution of benefits from fact-checking. 

To answer these questions, we develop the  Topic-Aware, Community-Impacted Twitter (TACIT) simulator, a novel agent-based methodology for simulating the spread and consumption of multi-topic information across networks, detailed in Section \ref{sec:approach}. We use the ``construct'' model of social influence and reasoning \citep{carley2009etiology} to guide probabilistic information consumption and propagation behaviors for agents in the simulation, and we build upon other recent simulation efforts \citep{beskow2019agent} to consider a multi-topic information environment in which communities are particularly impacted by certain topics of information. We run this simulation on a Twitter followers dataset with a strong community structure \citep{de2013anatomy}, and we tune the parameters of our simulation so it exhibits the emergence of information cascades of similar depth, width, and overall engagement to those derived from prior analysis of how true and false information spreads online \citep{vosoughi2018spread}. In Section \ref{sec:experiment}, we assess the network-wide and group-level efficacy of different check-worthiness algorithms that explicitly account for relevant notions of diversity. We find that both representative \textit{and} egalitarian methods for sampling and labeling training data can lead to network-wide benefit concentrated in majority communities, while incorporating diversity into how fact-checkers use algorithmic recommendations actively reduces inequalities in benefits between majority and minority communities.

%% file: related_work.tex
\subsection{Theoretical Models of Information Sharing and Belief}

Early efforts to explain information sharing and belief behavior sought to incorporate insights from research about the social nature of infection spread \citep{goffman1966mathematical, goffman1964generalization, daley1965stochastic, maki1973mathematical}. These models (often referred as SIR, or ``Susceptible, Infected, Recovered'') generally categorize agents into three buckets: ignorant (akin to ``susceptibles''), spreaders (``infected''), and stiflers (``recovered''). Criticisms of this research model have noted that anti-misinformation or anti-rumor is disseminated in rumor-like cascades \citep{TripathyRudra2010Asor}, which is a major dissimilarity to epidemic processes.  Additionally, while these early models are stochastic, they tend to assume away complex heterogeneity present in real world data. 

``Construct'' theory \citep{carley1991theory, carley2009etiology} provides a flexible, yet theoretically grounded, model for observed human behavior on social networks, drawing on research from the social and network sciences. Fundamentally, the theory states that humans within a social network possess bounded rationality \citep{simon1990bounded}. Bounded rationality implies that humans cannot recall or process all information available to them, and that the positioning of humans within a social context limits (or otherwise biases) their exposure to all available information. This theory has been used to explain and model how networks evolve over time \citep{carley2012dynamic}, how behavior mirroring influences consumer decisions \citep{gloor2017mirroring}, methods for de-escalating crises in various online communities \citep{lanham2013social}, the effect of malicious bots on information sharing behavior online \citep{beskow2019agent, benigni2019bot}, and the effects of quality management on organizational productivity \citep{jamshidnezhad2015agent}. By incorporating the propositions of bounded rationality into an agent-based model (ABM), we can better model the emergence of ``irrational'' phenomena on the internet and in the real world, such as the adamant belief in false information shared online. For instance, research on the ``illusory truth effect'', which shows that repeated encounters with fake news increase an individual's perception of its veracity \citep{hassan2021effects} and that prior knowledge is not sufficient to combat this effect \citep{fazio2015knowledge}, can be explained by a human's informational processing constraints (memory) and position within a social network (repetition). To date, this theory and research model have not been used to study the responsible use of fact-checking algorithms on social networks.

\subsection{Analyzing The Spread of Misinformation Online}

It comes as no surprise to anyone living in this age that misinformation spreads far and wide on the internet. Nonetheless, \citet{vosoughi2018spread}, in their seminal analysis of 128,000 rumor cascades spread by around 3 million people on social media, highlight the incredible extent to which falsehoods are more viral than truth online. They find significantly different distributions for cascades of true and false information, whereby false claims tend to have deeper cascades, wider cascades at any depth, a larger number of overall users engaged, and a higher structural virality \citep{goel2016structural} - a measure that interpolates between the depth of the rumor and its pervasiveness across a population. However, they do find that the right-tail of the distribution of true claims (i.e. the top 1\% most shared true claims) are shared more frequently, thus producing more cascades per claim than false information. 

Given the massive amount of information shared online and the proprietary nature of algorithms developed by social media platforms, agent-based modeling (ABM) can be a helpful tool for investigating research questions related to complex human behavior on social networks \citep{lanham2013social, bonabeauABM, jamshidnezhad2015agent} and the potential impacts of algorithmic interventions on complex systems \citep{simulation_foster_care}. We extend the simulation approach in \citet{beskow2019agent}, in which the authors simulated the spread of true and false information online in the presence of strategic bots, after confirming the simulation generates information cascades similar to those found in \citet{vosoughi2018spread}.



\subsection{AI-Assisted Fact-Checking: Check-worthiness Estimation}

\label{sec:checkworthines_related_work}

Recent evidence shows that most engagement with misinformation happens early in its lifespan, usually within the first 30 hours \citep{goldstein2023understanding}. Fact-checking organizations \citep{fullfact} and platforms \citep{meta_for_media_2021} are increasingly relying on algorithms to quickly identify claims circulating online that should be prioritized for fact-checking based on the linguistic and network-propagation patterns of the claim. This AI framework is typically called ``check-worthiness'' estimation. According to \citet{claimbuster} and \citet{snopes_2021}, check-worthy claims are claims for which the ``general public'' would be interested in assessing their validity. 

Predicting which claims to fact-check is a complex task given the population-level scale of modern social media. Several model architectures, information sampling mechanisms, and labeling paradigms have been proposed for detecting check-worthy claims circulating on social media (e.g.~\citep{checkthat, automated_claim_detection, claimbuster, twitter_metadata, checkworthy_multitask, hassan2019examining}). For example, in \citet{claimbuster}, the researchers gather numerous political experts and have them categorize sentences from political speeches as either \emph{Non-Factual Sentence}, \emph{Unimportant Factual Sentence}, or \emph{Check-Worthy Factual Sentence}. These labels are then used to train or fine-tune an algorithm for predicting future check-worthy claims, presumably made by politicians in speeches. In contrast, platforms such as Twitter have taken a crowd-sourcing approach. Rather than relying on experts to identify the check-worthiness of claims, Twitter asked everyday users on the platform to identify potentially misleading tweets \citep{allen2022birdwatch, SaeedCrowdsource22}. Additionally, \citet{checkworthy_multitask} source a number of highly viral claims about COVID-19 circulating on Twitter and investigate how a claim may be check-worthy in different regards (e.g. for society at large, fact checkers, vulnerable communities), finding that there is substantial variation in estimated check-worthiness depending on how the data labeling question is asked.

%% file: approach_edits.tex
Our research goal is to investigate how the benefits of AI-assisted fact-checking are distributed amongst different online communities present in a social network. Insights from this work could shed light on systemic deviations in access to quality information brought about by the use of these algorithms. This topic is difficult to study with existing datasets due to the opacity of decisions made by algorithm designers. For instance, borrowing from a recent framework by \citet{fazelpour2022diversity} for considering diversity in sociotechnical systems, any representation of the data learned by the AI tool will be dependent upon (1) the data sampled for model training, (2) who evaluated the ``check-worthiness'' of this training data (used as training labels), and (3) how these predictions were integrated into the fact-checker's workflow. Therefore, in Section \ref{sec:tacit}, we propose the Topic-Aware Community-Impacted Twitter (TACIT) simulator: a novel agent-based methodology for simulating the spread and consumption of multi-topic information across networks. Our simulation approach is able to generate behavior on networks with varying structures and ``states of the world'' while retaining realistic information propagation and consumption characteristics found in prior work \citep{goel2016structural, vosoughi2018spread}. Given its realistic data generating process, TACIT serves as a transparent test bed for check-worthiness algorithms. In Section \ref{sec:approach_models} we describe our approach to testing how different notions of diversity can be incorporated into algorithmic interventions designed to stop the spread of misinformation. 


\subsection{Topic-Aware, Community-Impacted Twitter (TACIT) Simulator}
\label{sec:tacit}
We extend the approach introduced in \citet{beskow2019agent}, which uses the construct model from \citet{carley2009etiology} to simulate information propagation and consumption on a Twitter-like micro-blogging platform. 

\subsubsection{Nodes, Edges, \& Communities}
TACIT takes as input a graph $\mathcal{G}$, comprised of a set of nodes $\mathcal{J}$ and directed edges $\mathcal{E}$. Nodes represent users on a micro-blogging platform, and they have the ability to tweet (create) information and simultaneously pass it on to their followers who then have the option to consume and further re-tweet (propagate) that information to their followers. The presence of a directed edge $(j^{[1]},j^{[2]})$ means that node $j^{[1]}$ follows node $j^{[2]}$, and would receive information tweeted and re-tweeted by node $j^{[2]}$. 

Each node in the graph is assumed to belong to a community $c \in \{0, 1, 2,...C\}$. We use $\mathcal{J}^c$ to denote the set of nodes that belong to community $c$. The community structure can represent ground-truth communities associated with the graph, or it can be mined from the graph itself using a community detection algorithm such as the Louvain method \citep{Blondel_2008}.

\subsubsection{Topics, Claims, \& Utterances}
\label{sec:claims_utterance}
The TACIT simulator generates a world in which there exist topics $\omega \in \Omega$, where $\Omega$ is the full set of possible topics. For each topic, there exist claims $\kappa^{\omega} \in K^{\omega}$, where $K^{\omega}$ is the full set of possible claims that can be made about a topic $\omega$. A claim $\kappa^{\omega}$ can be viewed as an assertion about the topic $\omega$. When an individual node $j$ tweets about a claim $\kappa^\omega$ at any time $t \in T$, they generate an utterance $\mu(j, \kappa^{\omega}, t) \in \{\mathcal{J} \times K^\omega \times T\}$. We use the veracity function $V(\kappa^\omega) \in \{-1,0, 1\}$ to denote the ``veracity'' of claims, where the values $\{-1,0, 1\}$ correspond to anti-misinformation, noise, and misinformation, respectively\footnote{Anti-misinformation claims contain ground-truth knowledge about a topic, while noise claims may only partially address a given topic.}. The veracity of an utterance $\mu(j, \kappa^\omega, t)$ is always equal to the veracity of its parent claim $\kappa^\omega$ such that $V(\kappa^\omega) = V(\mu(j, \kappa^\omega, t))$. To relate to previous work \citep{goel2016structural, vosoughi2018spread}, utterances are the start of potential information cascades in the network. 

Claims within a topic have inherently different levels of virality. Highly viral claims are more likely to become utterances (be tweeted) and to be re-tweeted. The virality of a claim $\kappa^\omega$ is defined by function $f_{v}(\kappa^\omega)$, shown in Eq. \ref{eq:claim_viral}, and the probability with which a node $j$ tweets (generates an utterance) at time $t$ about a claim $\kappa^\omega$ with $V(\kappa^\omega) = V$ is shown in Eq. \ref{eq:utterance_viral}:

\begin{equation}\label{eq:claim_viral}
f_{v}(\kappa^{\omega}) =
    \begin{cases}
        z_1+s_1, \text{where }s_1 \sim Beta(\alpha_1, \beta_1) & \text{if } V(\kappa^{\omega}) \in \{-1, 0\}\\
        z_2+s_2, \text{where }s_2 \sim Beta(\alpha_2, \beta_2) & \text{if } V(\kappa^{\omega}) = 1
    \end{cases}
\end{equation}

\begin{equation}\label{eq:utterance_viral}
P(K_V^\omega = \kappa^\omega | j, t) =
    \begin{cases}
        softmax(r_1*\vec{f_{v}}(K^\omega_{V})^{q_1}) & \text{ if } V \in \{-1, 0\}\\
        softmax(r_2*\vec{f_{v}}(K^\omega_V)^{q_2}) & \text{ if } V = 1
    \end{cases}
\end{equation}

In the above equations, $K^\omega_V$ denotes the set of all possible claims with topic $\omega$ and $V(\kappa^\omega) = V$, and $\vec{f_{v}}(K_V^\omega)$ denotes the vector of claim virality values for all claims in $K_V^\omega$. Therefore, the set of parameters $\alpha_1, \alpha_2, \beta_1, \beta_2, z_1, z_2, r_1, r_2, q_1, q_2$ dictate the size, depth, and breadth of information cascades for particular claims. In our experiments, we chose these parameters so that the information distributions produced by the simulation align with current empirical research \citep{vosoughi2018spread}.

\begin{figure*}
 \centering
    \includegraphics[width=0.7\textwidth]{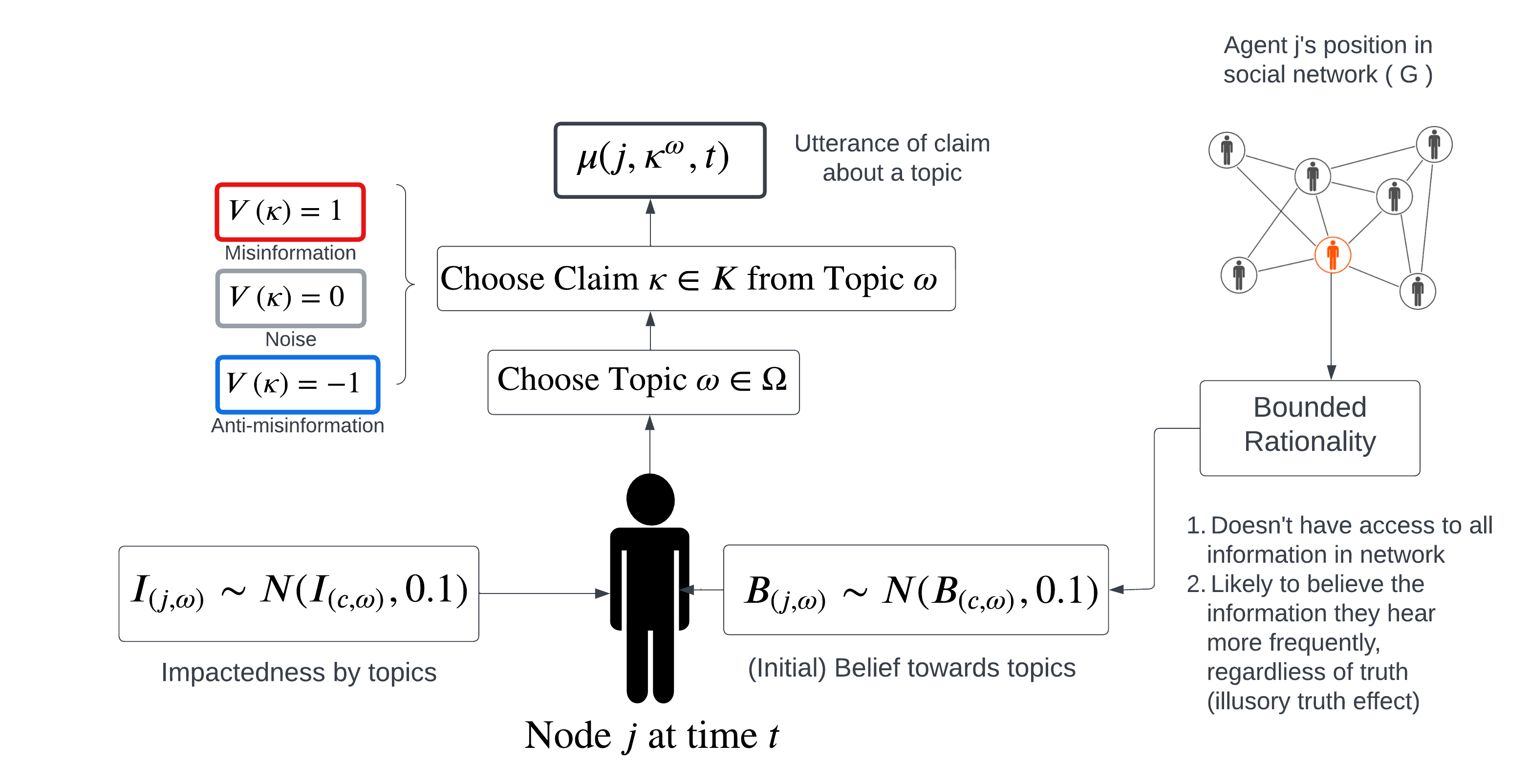}
    \caption{\textbf{Information propagation in TACIT.} Belief and Impactedness towards topics are initialized dependent on community membership, and belief is subsequently updated depending on the veracity of information to which a node is exposed (Eq. \ref{eq:belief_update}). Nodes generate an utterance about a claim, dependent on their belief and impactedness towards topics, and the utterance is propagated to their followers. Changes in belief create new propagation patterns for nodes. We assume impactedness towards a topic is fixed.}
    \label{fig:tacit_info_prop}
\end{figure*}

\subsubsection{A Node's Relationship to Topics} 
\label{sec:impact_belief}

As demonstrated in Figure \ref{fig:tacit_info_prop}, we focus on two key properties that dictate: (i) a node's propensity to produce true and false information on a given topic; (ii) their interest in talking and reading about various topics. We define these constructs as ``Belief'' and ``Impactedness'' respectively. Belief is borrowed from the ``construct'' theory and model \citep{carley1991theory, carley2009etiology} and reflects a node's ground truth understanding of the world, such that a higher belief represents a lesser understanding. A node's belief changes over time, and is influenced by (i) their randomly assigned starting belief, and (ii) the information they are exposed to in the network. Having a high belief towards a topic means that a node is more likely to consume, produce, and propagate misinformation about that topic. Similarly, nodes with low belief are more likely to consume, produce, and propagate anti-misinformation (truth or knowledge). 

A node's impactedness towards a topic reflects the importance of the topic to the node. Nodes will tend to tweet and retweet claims about high-impact topics more frequently in TACIT. Impactedness can vary amongst communities for a given topic, or be largely constant. For instance, a topic like the outbreak of an epidemic would likely have high impact across all online communities, while the topic of immigration law may be particularly important to a community of non-citizen workers.

Further, we assume that nodes organize in communities due to their shared world view, which means that nodes in the same online community will have highly correlated belief and impactedness values towards topics \citep{echo_chamber_effect}. Therefore, each \emph{(community, topic)} pair is assigned an impactedness $I_{(c,\omega)} \in [0,1]$ and belief $B_{(c, \omega)} \in [0,1]$ at the start of the simulation.  Node $j$ is assigned impactedness and belief values $I_{(j,\omega)}$ and $B_{(j, \omega)}$ respectively, which are drawn from Gaussian distributions with $I_{(c,\omega)}$ and $B_{(c, \omega)}$ as the mean. These parameters represent the ``state of the world'' for nodes in the various communities in $\mathcal{G}$, and by carefully choosing these parameters, one can examine many scenarios that can be observed across real-world social networks.

Belief is not static in TACIT, and, as the ``illusory truth effect'' \citep{hassan2021effects, fazio2015knowledge} implies, repeated exposure to misinformation makes a node more likely to believe it to be true, regardless of their starting belief. Drawing from the cognitive psychology literature \citep{hogarth1992order}, we assume that a node $j$'s belief is updated upon viewing their $n^{th}$ utterance about topic $\omega$ as follows:

\begin{equation}\label{eq:belief_update}
   B_{(j, \omega)}:= B_{(j,\omega)} + \alpha*\frac{1}{n+1}V(\mu(\cdot, \kappa^\omega, \cdot))\big(1+I_{(j, \omega)}\big)
\end{equation}

where $\alpha$ is a learning rate parameter. Intuitively, Eq. \ref{eq:belief_update} iteratively updates belief by the value of the information $V(\mu(\cdot, \kappa^\omega, \cdot))$ scaled by the impactedness of the node $I_{(j, \omega)}$. Misinformation increases belief, anti-misinformation decreases belief, and noise has no effect on belief.

\subsubsection{Types of Nodes}
``Normal'' nodes make up the majority of nodes in the network and represent the average micro-blogger. A normal node $j$ generally creates utterances about claims $\kappa^\omega$ where $V(\kappa^\omega) = 0$ (are noise) in the simulation, with the chance of producing utterances about claims with either $V(\kappa^\omega) = -1$ or $V(\kappa^\omega) = 1$, depending on their beliefs towards topic $\omega$, such that nodes with relatively higher belief regarding a topic have a higher chance of producing misinformation, while those with lower belief have a higher chance of producing anti-misinformation. A percentage of nodes are randomly selected to be ``bots''. Bots are malicious, and generate misinformation at a higher rate than any normal node. 

\subsubsection{Simulation Procedure} 
The pseudocode for the simulation procedure is shown in Algorithm \ref{alg:TACIT}. The simulation takes as input a graph $\mathcal{G}$ and runs for $T$ time steps. At each time step $t$, the simulation iterates through all the nodes in the graph. A node first has an opportunity to tweet. We assume that if the node is a bot, they will tweet every time they are active \citep{beskow2019agent}. For normal nodes, we assume that their tweeting is proportional to their popularity in the network, such that more popular (higher ``prestige'' in Algorithm \ref{alg:TACIT}) nodes will tweet more frequently \citep{huberman2008social}. 

If a node decides to tweet, they first select the topic $\omega$ to tweet about dependent upon $I_{(j,\omega)}$ over the set of topics $\Omega$, such that they will tweet about topics which impact them more. $B_{(j, \omega)}$ informs the veracity of their utterance about a topic, such that those with low belief will produce ``true'' utterances more frequently and those with high belief will generate misinformation more frequently. Finally, node $j$ selects a claim from the set of claims $K^\omega_V$ according to the probability distribution shown in Eq. \ref{eq:utterance_viral}. After a node has its opportunity to tweet, the node reads a random subset of their inbox. Each utterance the node reads updates the node's belief about the utterance's topic according to Eq. \ref{eq:belief_update}. Next, the node has a chance to further retweet the utterance, again propagating it to all of their followers (adding it to their inbox). The probability with which node $j$ re-tweets an utterance $\mu(i^{*}, \kappa^\omega, t)$ that originated with node $i^*$ is shown in Eq. \ref{eq:retweet}:

\begin{equation}
\label{eq:retweet}
P(j\text{ re-tweets }\mu(i^{*}, \kappa^\omega, t)) \propto
    k^{in}_{i^*}*f_{v}(\kappa^{\omega})*\begin{cases}
        (1-B_{(j, \omega)}) & \text{if } V(\kappa^\omega) = -1\\
        0.5 & \text{if } V(\kappa^\omega) = 0\\
        (B_{(j, \omega)}) & \text{if } V(\kappa^\omega) = 1\\
    \end{cases}
\end{equation}

Where $k^{in}_{i^*}$ is the number of followers node $i^*$ has. Eq. \ref{eq:retweet} implies that the probability with which node $j$ re-tweets an utterance increases as the prestige of node $i^*$ increases, and as the virality of the underlying claim increases. If the utterance is misinformation, a higher belief in the utterance's topic increases the probability of re-tweeting, while if the utterance is anti-misinformation, a higher belief in the utterance's topic decreases the probability of re-tweeting. The node's belief has no impact on the likelihood that it re-tweets noise. Finally, after reading tweets, the node clears their inbox, and the next node goes through the information creation and propagation phase.

\begin{algorithm}[t]
\caption{TACIT Simulator}\label{alg:TACIT}
\begin{algorithmic}[1]
\State \textbf{Input:} $G, T$
\For{$t=0,1,...T$}
\For{node $j$ in $G$}
\If{$j$.Wake}
\If{$j$.Kind == `bot' $||$ $j$.Prestige $\geq$ random()}
\State $j$.create\_tweet() \Comment{create and append new utterance to all followers of $j$ (see figure \ref{fig:tacit_info_prop})}
\EndIf
\For{utterance $\mu$ in subset($j$.Inbox)}
\State Update $j$.Belief \Comment{Update belief using update in Eq. \ref{eq:belief_update}}
\If{j.retweet\_perc($\mu$) $\geq$ random()}
\State $j$.re-tweet($\mu$) \Comment{append utterance to all followers of $j$}
\EndIf
\EndFor
\EndIf
\State Clear $j$.Inbox
\EndFor
\EndFor
\end{algorithmic}
\end{algorithm}

\subsection{The Role of Diverse Representation in AI-Assisted Fact-Checking}
\label{sec:approach_models}

When considering check-worthiness models, we choose not to focus on the impact of model architecture and training data features, so we use a single out-of-the-box gradient-boosted tree regressor \citep{xgboost} and the same set of features across all experiments. Instead, we employ a framework from \citet{fazelpour2022diversity} to investigate how diverse representation within training data and human-AI collaboration impacts the distribution of benefits from algorithm-assisted fact-checking across communities. In particular, we vary: (i) how claims are sampled from the network to produce a training dataset; (ii) how and from whom opinion labels used to train the algorithm are queried, and (iii) how these predictions are ultimately used to inform downstream fact-checking.

In our simulation, a check-worthiness model intervention is implemented at time step $T_m$ using data gathered from time steps $0,...,T_m$. At each time step $t = 0,...,T_m$, for each claim $\kappa$ across all topics, we tabulate a set of features $x_{\kappa}$. The features characterize the average depth, width, and overall spread of utterances about claim $\kappa$ as it passes through a network at various time-from-creation snapshots. Additionally, $x_{\kappa}$ captures information about the nodes propagating and creating the information such as average degree and betweenness centrality. We intend only to create features that would be available to real-world platforms, and as such, we do not capture anything related to the impactedness and beliefs that nodes might have towards claims they are spreading\footnote{For a precise listing of features, please see appendix Section \ref{sec:appendix_features}}. The data are then used to train a check-worthiness model which will be implemented for time steps $T_m,...,T$, where features are continuously updated, and claims that were fact-checked at time $t^{*} < t$ will be excluded from the prediction set.

After time $T_{m}$, if a particular claim is predicted to be check-worthy, it is sent to fact-checkers in that same time step. If this claim is determined to be misinformation, it can no longer be posted or read again by any node in the network. This mitigation approach prevents the spread of fact-checked misinformation entirely, allowing us to directly assess the \emph{relative} impact of the various interventions under consideration. 

\subsubsection{Diversity Variable 1: How are Claims Sampled?}

In practice, it is simply too expensive to acquire check-worthiness training labels for all but a small subset of claims passing through highly populated social networks due to the scale at which information is shared. Therefore, given the constraint that only $n$ claims can be selected for model training, how should these $n$ claims be selected? A baseline approach that does not consider community structure in networks is to select the $n$ claims whose utterances have produced the largest cascades across the network (i.e. \textbf{virality sampling}). This approach is computationally more straightforward and employed by researchers \citep{checkworthy_multitask, claimbuster} and fact-checking organizations \citep{snopes_2021, fullfact} alike. However, we are concerned that it may introduce bias into the algorithm's representation of check-worthiness as claims circulating in minority communities in the network are unlikely to be sampled in the training dataset. To remedy this issue, we introduce a \textbf{stratified virality sampling} approach, which selects the $n/C$ claims that have been tweeted and retweeted the most in each community $c \in 1,..,C$. This method should ensure that an equal number of claims that have gone viral within each community are represented in the check-worthiness training dataset, and as such, evokes an egalitarian notion of diversity \cite[pg. 3]{fazelpour2022diversity}.


\subsubsection{Diversity Variable 2: Who Labels the Check-Worthiness of Claims?}

Once a subset of $n$ claims is selected, each selected claim $i$ needs to be labeled, producing a $n \times 1$ label vector $Y$. The label $y_i \in [0,1]$ of a claim $i$ represents how check-worthy the claim is perceived to be by a selected sample of the network population. The check-worthiness of a claim is a subjective value unique to each individual. Let the label for claim $i$ provided by labeler $j$ be denoted by $y^{[j]}_i \in \{0,1\}$, where $0$ means $j$ believes claim $i$ to not be check-worthy, and $1$ means $j$ believes it to be check-worthy. For each approach, the final label for each claim $i$ can be written as $y_i = \frac{\sum_{j=0}^m y^{[j]}_i}{m}$.

The final label $y_i$ is the proportion of all $m$ labelers that identified the claim to be check-worthy. In the TACIT simulation, each labeler $j$'s labeling decision is guided by their belief towards claim $i$'s topic $\omega$ and can be described as drawn from the distribution shown in Eq. \ref{eq:node_label}:

\begin{equation}\label{eq:node_label}
y^{[j]}_{i^\omega} \sim 
    \begin{cases}
        Bernoulli(1-B_{(j,\omega)}) & \text{if } V(i) = -1\\
        Bernoulli(0.05) & \text{if } V(i) = 0\\
        Bernoulli(B_{(j,\omega)}) & \text{if } V(i) = 1\\
    \end{cases}
\end{equation}

If the claim is anti-misinformation, the likelihood that node $j$ labels the claim as check-worthy is \emph{inversely proportional} to their belief about the claim's topic. If the claim is misinformation, the likelihood that node $j$ labels the claim as check-worthy is \emph{proportional} to their belief about the claim's topic. Finally, if the claim is noise, there is a small chance that the node labels it as check-worthy, regardless of their belief. 

Given that a claim's check-worthiness is the average of a set of queried opinions, the claims which are identified as check-worthy are highly dependent on the labelers selected to label the claims. Some online communities contain highly specialized and accurate knowledge, and members of these communities should be considered as a highly credible - and economical - resource for knowledge management in social networks \citep{wang2013expertrank}. Recent work on the ethics of fact-checking systems states that ignoring knowledgeable ``sources of evidence'' can result in an ``epistemic injustice'' for those whose lived experiences provide valuable evidence towards claims \cite[pg.1507]{neumann2022justice}.

To investigate how the choice of labelers affects an algorithm's learned representation of check-worthiness, we introduce several claim labeling approaches: \textbf{(i) Random Labeling} - This captures \emph{representative} diversity \cite[pg. 3]{fazelpour2022diversity} in the system, such that the proportions of the different communities reflect their expected contribution to the label. For each claim $i$, $m$ nodes are randomly selected from the entire network (i.e. community structure is ignored), and their opinions towards $i$ are averaged;  \textbf{(ii) Stratified Labeling} - This captures an \emph{egalitarian} notion of diversity \cite[pg. 3]{fazelpour2022diversity}, such that each group is equally represented. For each claim $i$, $m/C$ labelers are selected from each community $c \in 1,..,C$, and their opinions towards $i$ are averaged; and \textbf{ (iii) Knowledgeable Community Labeling} - A community known to be knowledgeable about a topic is surveyed for their check-worthiness opinions. For each claim $i$ about topic $\omega$, $m$ labelers are selected from community $c$ with lowest average belief $B_{(c,\omega)}$ towards topic $\omega$, and their opinions are averaged.

\subsubsection{Diversity Variable 3: How are Predictions Used by Fact-Checkers?}
Finally, we consider how model predictions influence fact-checkers' decision-making processes about which set of claims to fact check. Let $Z^*$ denote the set of all claims that have been selected for fact-checking up until time step $t$, and let $X_t$ denote the matrix of features of all claims that have been tweeted and re-tweeted during time $t$, but are not in $Z^*$. The predicted values $f(X_t)$ are utilized to select $z$ claims for fact-checking. We introduce two methods for utilizing the output of $f(X_t)$ to select $z$ claims to fact-check per time step: \textbf{(i) Check the Top Predicted} - The claims with the top $z$ values in $f(X_t)$ are selected for fact checking and added to set $Z^*$; and \textbf{(ii) Check the Top Predicted by Topic} - The top $z/|\Omega|$ predicted claims for each topic $\omega \in \Omega$ are selected for fact checking and added to set $Z^*$.

\subsubsection{Model Training Procedure}

We use out-of-the-box XGBoost regressors \citep{xgboost} in the model training phase, and we train two separate models. The first model learns a regressor $f_1(X) \rightarrow Y$, where $X$ is the features matrix and $Y$ is the check-worthy label vector. This regressor is used to predict the check-worthiness of all claims made across $\mathcal{G}$. We also train a second regressor $f_2(X) \rightarrow S$, where $S$ contains values $s_i$ for all claims $i$, and where $s_i$ is the average number of nodes that read and/or interacted with claim $i$ per utterance. Regressor $f_2$ is thus used to predict the virality of claim $i$. Next, for any claim $i$, we generate a final check-worthiness predicted value $f(x_i) = f_1(x_i)*f_2(x_i)$. Thus, a claim's predicted value is proportional to both the predicted perceived check-worthiness of the claim and the predicted virality of the claim.

%% file: experiment.tex
\begin{figure*}
 \centering
    \includegraphics[width=0.9\textwidth]{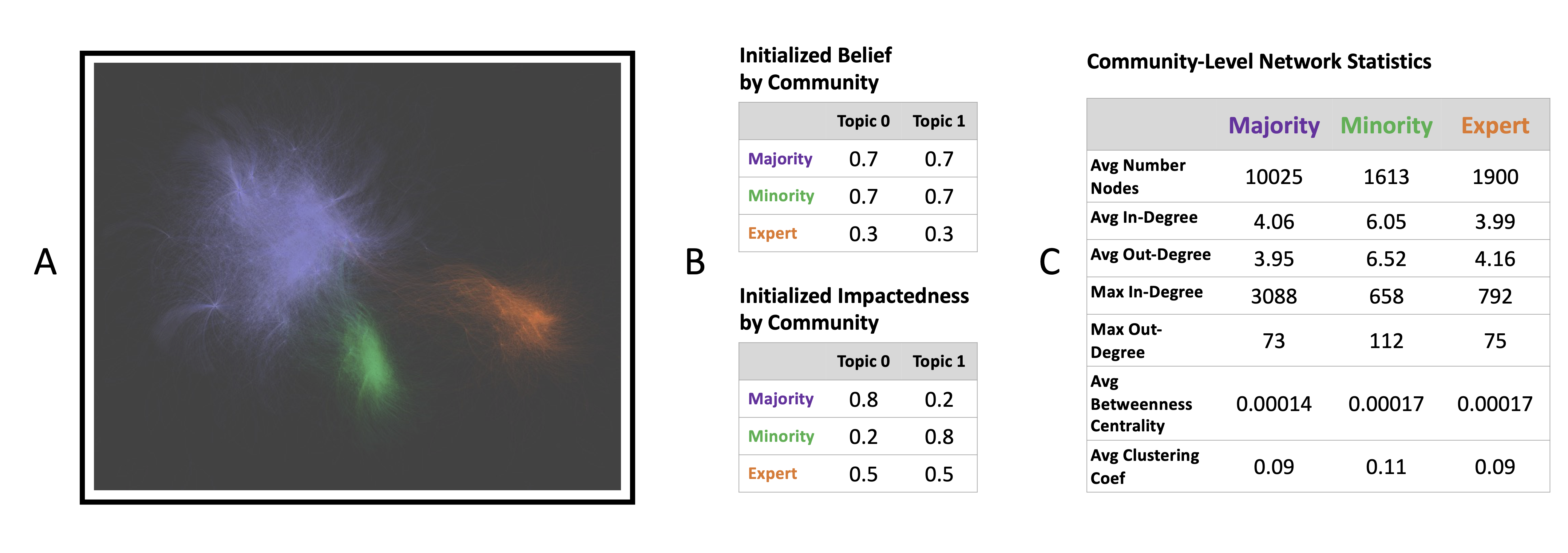}
    \caption{\textbf{A: Network Visualization.} This figure shows a 15\% sample ($|\mathcal{J}| = 13,538$) of three communities from the network first published in \citet{de2013anatomy}. \textbf{B: Simulation Parameters.} These tables show the initialized values for belief and impactedness towards topics. Node-level values will be drawn from Gaussian distributions, with these values as the mean and a constant standard deviation. \textbf{C: Community-Level Statistics.} Differences in average node-level characteristics across the three communities. The majority community consists of approx. 75\% of the nodes in the network.}
    \label{fig:community_structure}
\end{figure*}

In order to investigate our research question, we run TACIT on a Twitter followers network dataset with a strong (albeit not ground-truth) community structure\footnote{This community structure was detected using standard settings of the Louvain method in \texttt{NetworkX}. It is likely these different communities, given the context of \citet{de2013anatomy}, represent geographically diverse Twitter users interested in scientific advancement.} that accounts for roughly 75\% of the 450,000 Twitter users first analyzed in \citet{de2013anatomy}. From this dataset, we focus specifically on three communities: a large majority community and two smaller minority communities, displayed in Figure \ref{fig:community_structure}A, with average community statistics in Figure \ref{fig:community_structure}C.

We investigate a world in which there exists a \emph{majority} community, a \emph{minority} community, and a small \emph{expert} community \citep{experts_on_twitter, twitter_demogs}. We further assume that $|\Omega| = 2$, with the majority community being more impacted by $\omega_{0}$ and the minority community being more impacted by $\omega_{1}$. Additionally, in this world, both majority and minority communities are equally un-knowledgeable about all $\omega \in \Omega$. The only community with consistent ground-truth knowledge about the topics is the small expert community, who is equally impacted by both topics. This is reflected by the parameter settings shown in Figure \ref{fig:community_structure}B. While the communities and topics are abstract values within the simulation, they represent possible real-world scenarios. For example, imagine that the network represents Twitter users that are interested in labor law in the United States, and that the majority community are mostly U.S. citizens while the minority community are mostly immigrants working in the U.S. $\omega_{0}$ can represent claims about labor laws pertaining to U.S. citizens (e.g. rumored changes to collective bargaining rights for workers of large U.S.-based corporations), while $\omega_{1}$ can represent claims about immigrant labor laws (e.g. rumored changes to H-2B work permits.) These topics should impact the majority and minority communities in our network differently. Finally, the expert community in this context could be U.S. lawyers and lawmakers working on labor-related issues.

\begin{figure*}
 \centering
    \includegraphics[width=\textwidth]{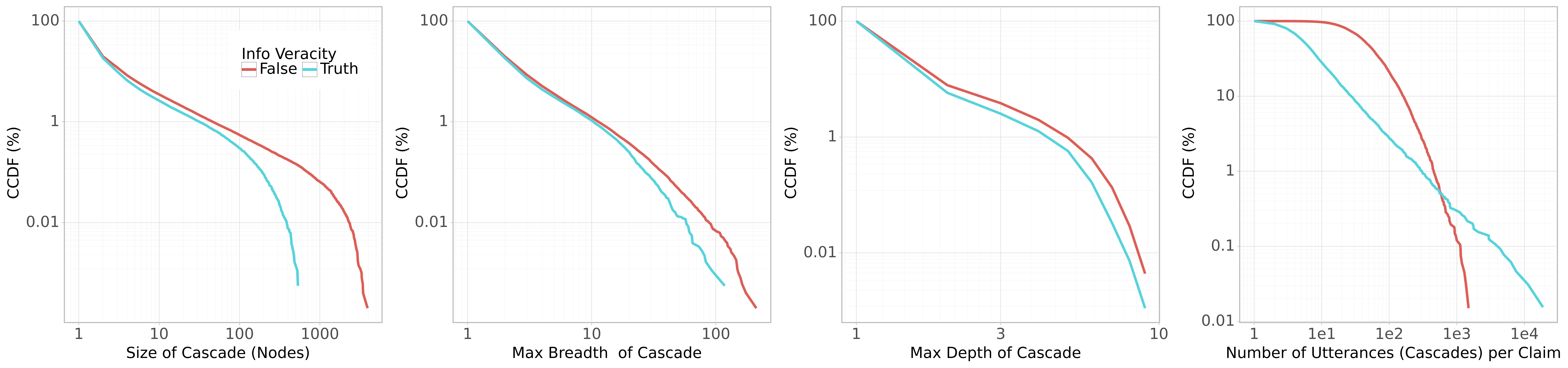}
    \caption{\textbf{Simulation Validation.} These complementary cumulative distribution functions (CCDFs) demonstrate differences in information diffusion by veracity in our simulation, similar to a seminal analysis conducted by \citet{vosoughi2018spread}. We see that once the parameters in Eqs. \ref{eq:claim_viral} and \ref{eq:utterance_viral} are properly tuned, our simulation follows largely the same patterns established in the empirical literature. Falsehoods travel deeper, wider, and are seen by more nodes than truth. The last subplot, also consistent with data, shows that some true claims (e.g. election results) generate significantly more utterances than any false claim.}
    \label{fig:ccdf}
\end{figure*}

\subsection{Experimental Approach}

\subsubsection{Validating and Tuning TACIT}

In order to make sure that our information cascades are similar to those observed on real social networks, we tune the parameters in Eqs. \ref{eq:claim_viral} \& \ref{eq:utterance_viral} and compare complementary cumulative distribution functions (CCDFs) produced by TACIT of (i) cascade depth, breadth, and size, and (ii) the number of utterances per claim to the same CCDFs presented in \citet{vosoughi2018spread}. We find that setting $z_1 = 1$, $z_2 = 1.5$, $s1 \sim Beta(1, 17)$, and $s2 \sim Beta(5, 17)$ produces the desired behavior in claim virality such that, across the entire distribution, false utterances travel deeper, wider, and are seen by more people than true utterances. By setting $r1 = 9$, $r2 = 9$, $q1 = 2$, and $q2 = 1$, we produce the desired distributional characteristics of utterances per claim; true claims in the far right tail will have more generated utterances across the network than false claims at the far right tail. Intuitively, this implies that some facts are unambiguously true and important, and they receive more attention than any misinformation claim. See Figure \ref{fig:ccdf} for TACIT's CCDFs, and \citet[pg. 1147-1148]{vosoughi2018spread} for the comparison set of CCDFs.

\subsubsection{Outcomes \& Treatment Effects}

To answer our research questions, we use TACIT to observe the effects of different fact-checking approaches on network- and community-level outcomes. During the \textbf{pre-period} ($t < T_m$), information flows unmitigated through the network. Then, we take a snapshot of all nodes and their attributes at $T_m$ and apply all combinations of algorithmic components as described in section \ref{sec:approach_models} independently and in parallel in the \textbf{post-period} ($t \geq T_m$), including one run with no mitigation in the post-period. In order to better capture the impact of the algorithm across all topics for a given node, we introduce a measure called ``Impactedness-Weighted Change in Belief'' (IWCiB) such that $\text{IWCiB}_j = \sum_{\omega \in \Omega} \frac{I_{j,\omega}}{\sum_{\omega \in \Omega}I_{j,\omega}} * \big[B_{j, \omega}^{[t=T]} - B_{j, \omega}^{[t=T_m]} \big]$. $\text{IWCiB}_j$ represents $j$'s change in understanding of all topics from the midpoint to the end of the simulation, where greater weight is placed on topics that are important to the node. A negative IWCiB value for node $j$ means that $j$'s understanding of the world has improved from times $T_m$ to $T$. Note that while IWCiB is meant to represent a change in a node's understanding of the world, it is functionally a measure of how much anti-and mis-information a node was exposed to throughout the simulation, where greater weight is placed on high-impactedness information. As such, it can be used to assess the kind of misinformation, and quantity of misinformation, removed by different mitigation methods. 

Extending this approach to whole communities, the average treatment effect (ATE) for a mitigation $M_1$ on community $c$ can then be defined as $ \text{ATE}_{(M_1,c)} = \frac{1}{|\mathcal{J}^c|}\sum_{j \in \mathcal{J}^c}\big[\text{IWCiB}_{(M_1,j)} - \text{IWCiB}_{(M_0,j)}\big]$, where $\text{IWCiB}_{M_0, j}$ represents the resultant IWCiB for $j$ from no mitigation across the post-period, and $\text{IWCiB}_{(M_1,j)}$ is the IWCiB with mitigation $M_1$ implemented. This provides an accurate estimate for the average treatment effect for a given mitigation approach, because every node $j \in \mathcal{J}$ is observed in a state with various mitigations in place from time $T_m,...,T$ and, simultaneously, in a state without a mitigation ($M_0$) over the same period. We run 10 repetitions of the counterfactual experiment. For each repetition, we randomly subset 15\% of the total nodes from our three communities. Thus, every repetition will be an experiment on a different graph $\mathcal{G}$ with the same underlying community structure.

\subsection{Results}

\begin{figure}[t]
    \centering
    \includegraphics[width=\textwidth]{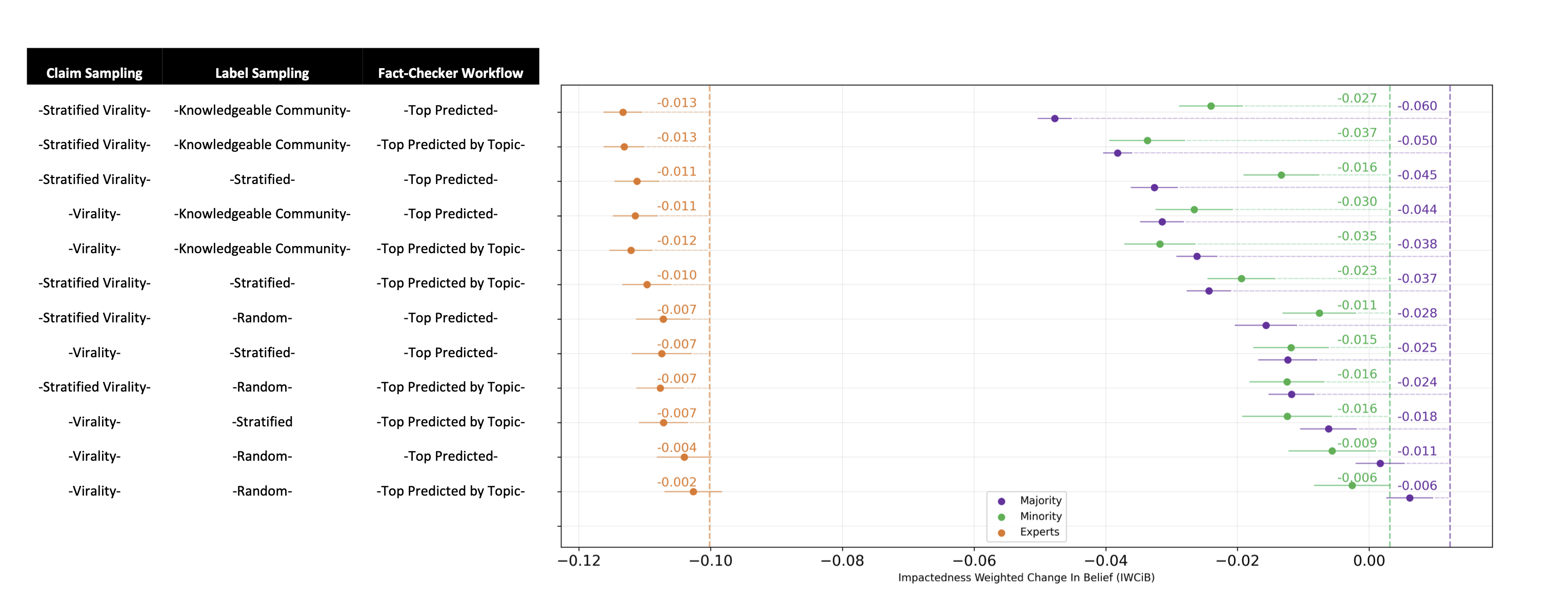}
    \caption{Average Impactedness Weighted Change in Belief (IWCiB) for all mitigation methods and communities. Points with solid horizontal lines show average IWCiB outcomes $\pm$ the standard error. Vertical dashed lines show IWCiB outcomes with no intervention for each community. Dashed horizontal lines and color-coded numbers  correspond to the Average Treatment Effects for all mitigation approaches on the three communities.  }
    \label{fig:megaPlot}
\end{figure}

We show the average treatment effect and final IWCiB outcomes (averaged across 10 repetitions) for all mitigations on the three communities in Figure \ref{fig:megaPlot}. In Table \ref{tab:result}, we show the average treatment effects resulting from the individual components of the mitigations, as well as the \emph{disparity ratio}, which is the ratio of the Majority-to-Minority ATE. A disparity ratio of 1 indicates that both the majority and minority benefit equally, while a disparity ratio greater than 1 indicates that the majority benefits more than the minority. 

The values in Figure \ref{fig:megaPlot} and Table \ref{tab:result} are useful \emph{descriptive statistics} for understanding outcomes in TACIT after the implementation of different AI-assisted fact-checking interventions. While we believe that we have developed a simulation with reasonable agent behavior that reproduces realistic distributions of information cascades, our findings do rely upon assumptions that some might disagree with. Therefore, these results are not \emph{inferential} insofar as they represent actual parameter values that might be observed in real data. Instead, we suggest that these results can help \emph{order and rank} the different effects observed, providing useful information to socio-technical theorists and policy makers about potential consequences of actions in a reasonably construed world. 

\paragraph{Does the Majority Community Disproportionately Benefit?}
As shown in Table \ref{tab:result} and Figure \ref{fig:megaPlot}, the majority community indeed disproportionately benefits under “representative” (Random Label Sampling and Virality Claim Sampling) \emph{and} “egalitarian” (Stratified Label Sampling and Stratified Virality Sampling) methods for assembling training data. Egalitarian methods for both claim sampling and label sampling lead to significantly higher network-wide benefit, but this benefit is \emph{more} concentrated in the majority community compared to representative diversity components. For example, in Table \ref{tab:result} we see a network-wide benefit of $-0.013$ from switching from Virality Sampling to Stratified Virality Sampling. However, after this switch, the minority community's outcomes improve only by $-0.004$, while the majority group's outcomes improve by $-0.017$, leading to an increased disparity ratio in treatment effects from $1.33$ to $1.86$. This provides suggestive evidence that, as the U.S. Senate \citep{congress2020} and European Union \citep{szakacs2021impact} sponsored research has pointed out, misinformation that impacts minority groups may be falling through the cracks at a higher rate than for majority groups online.

\paragraph{The (Equal) Value of Knowledgeable Online Communities}
There is an overwhelming benefit to using knowledgeable labelers to assess the perceived check-worthiness of claims. As shown in Table \ref{tab:result}, the Knowledgeable Community Labeling approach leads to an additional $-0.026$ and $-0.015$ network-wide benefit compared to random and stratified approaches. This network-wide increase in benefit is more equally distributed among the majority and minority communities with a treatment effect disparity ratio of 1.5 using Knowledgeable Community labelers compared to disparity ratios of 1.7 and 1.82 for Random Labels and Stratified Labels respectively. Therefore, the Knowledgeable Community Labeling approach significantly improves the outcomes across the network while also decreasing the relative disparity in treatment effects between communities compared to the other methods we tested. While this is an intuitive result, it demonstrates how certain \emph{communities of users} on a platform can perform a valuable service by providing their opinions, and it points to a way forward after other recent attempts at crowd-sourcing check-worthy claims were largely unsuccessful \citep{allen2022birdwatch}.

\paragraph{The Role of AI Predictions in Fact-Checker Workflow}

The data in Table \ref{tab:result} show that by switching from always fact-checking the top $n$ predicted claims (regardless of topic) to always fact-checking the top $n/|\Omega|$ predicted claims for each $\omega \in \Omega$, fact-checkers can \emph{actively reduce} the inequality in benefits from fact-checking. However, this comes at a cost to the network as a whole. From the results, switching from Top Predicted to Top Predicted by Topic \emph{decreases} the average treatment effect of the majority group by $+0.006$ and it \emph{increases} the minority average treatment effect by $-0.004$, leading to a drop in disparity ratio of treatment effects from $1.94$ to $1.32$ and a reduction in network-wide average treatment effect of $+0.005$. A key implication is that if fact-checkers can identify how different topics impact different online communities, they can control the distribution of benefits amongst these communities by choosing a specific level of topic representation in the set of claims they choose to fact-check. By choosing to fact-check topics relevant only to the majority community, you can still improve average outcomes across the network, but it will significantly increase the disparity in treatment for minority groups impacted by other topics that fact-checkers tend not to write about. If there are minority groups that are particularly at risk of being harmed by misinformation or who are particularly uninformed about topics, it is likely ethical and warranted to divert attention to fact-checking topics relevant to these communities.

\begin{table*}[t]
\resizebox{\textwidth}{!}{\begin{tabular}{ll|lll|l}
\hline
\multicolumn{2}{c|}{Mitigation} &
  \multicolumn{3}{c|}{Average Treatment Effect} &
  \multicolumn{1}{c}{\multirow{2}{*}{Disparity Ratio}} \\ \cline{1-5}
\multicolumn{1}{c|}{Diversity Variable} &
  \multicolumn{1}{c|}{Component} &
  \multicolumn{1}{c|}{Network} &
  \multicolumn{1}{c|}{Majority} &
  \multicolumn{1}{c|}{Minority} &
  \multicolumn{1}{c}{} \\ \hline \hline
\multicolumn{1}{l|}{\multirow{2}{*}{Claim Sampling}} & Virality                & -0.021 $\pm$ .001 & -0.024 $\pm$ .002 & -0.018 $\pm$ .001 & 1.33 \\
\multicolumn{1}{l|}{}                                & Stratified Virality     & -0.034 $\pm$ .002 & -0.041 $\pm$ .002 & -0.022 $\pm$ .001 & 1.86 \\ \hline
\multicolumn{1}{l|}{\multirow{3}{*}{Label Sampling}} & Random                  & -0.015 $\pm$ .001 & -0.017 $\pm$ .002 & -0.010 $\pm$ .001 & 1.7  \\
\multicolumn{1}{l|}{}                                & Stratified              & -0.026 $\pm$ .002 & -0.031 $\pm$ .002 & -0.017 $\pm$ .001 & 1.82 \\
\multicolumn{1}{l|}{}                                & Knowledgeable Community & -0.041 $\pm$ .002 & -0.048 $\pm$ .002 & -0.032 $\pm$ .001 & 1.5  \\ \hline
\multicolumn{1}{l|}{\multirow{2}{*}{Fact-Checker Workflow}} & Top Predicted           & -0.030 $\pm$ .002 & -0.035 $\pm$ .002 & -0.018 $\pm$ .001 & 1.94 \\
\multicolumn{1}{l|}{}                                & Top Predicted by Topic  & -0.025 $\pm$ .001 & -0.029 $\pm$ .002 & -0.022 $\pm$ .001 & 1.32 \\
\hline
\hline
\end{tabular}}
\caption{This table shows the ATE $\pm$ SE(ATE) for each component of the mitigation. Only comparisons within a diversity variable should be made. Claim sampling by stratified virality across communities leads to significant benefit concentrated in the majority community. Training on labels from experts leads to significant gains across communities. Fact-checkers stratifying their workflow by topic actively decreases inequality in the distribution of benefits.}
\label{tab:result}
\end{table*}


%% file: conclusion.tex
In this paper, we make several contributions to conversations around responsible use of AI-assisted fact-checking. First, we developed a novel simulator - TACIT - that reflects a world in which (i) people have bounded rationality; (ii) people form online communities based on shared interests; and (iii) people are more impacted by information from topics they find important. Our simulator can produce a wide-range of possible information consumption and propagation dynamics, and we were able to tune TACIT so that it produces realistic cascades of information identified in prior work. Second, we used a conceptual framework for diversity in socio-technical systems \citep{fazelpour2022diversity} to develop several check-worthiness model training and utilization approaches that incorporated varying forms of diverse representation. Third, we implemented our fact-checking approaches within TACIT to assess the impact of diverse representation on the distribution of benefits from AI-assisted fact-checking, finding that representative and egalitarian approaches to sampling and labeling training data can lead to network-wide benefit concentrated in majority communities. Further, we found that incorporating diversity into how fact-checkers use algorithmic recommendations can actively reduce inequalities in benefits between majority and minority communities. Finally, we make our code framework publicly accessible and encourage others interested in responsible AI-assisted fact-checking to utilize and expand on the simulation framework to further investigate questions related to the spread and detection of misinformation across online networks\footnote{To preserve anonymity in peer review, we will link to code at later date.}.

\subsection{Limitations}


In this paper, we abstract away the complex task of categorizing claims by their topics. In the real world, this likely adds a significant amount of noise to the efficacy of any check-worthiness algorithm, as it is yet another step in the pipeline where bias can exist. Future work on TACIT or related simulations should enable agents to generate text about certain topic (via generative technologies like ChatGPT, for instance) and evaluate model text classification approaches for topic modelling and textual check-worthiness estimation.

Finally, we do not investigate whether certain community-level network properties other than the size of communities can exacerbate or alleviate the differences in fact-checking outcomes, but this is likely the case. Future work should examine how community-level network characteristics impact the distribution of benefits.

\subsection{Implications}

Our findings show knowledgeable online communities can be an extremely valuable resource for training an algorithm to surface misinformation, providing sizeable benefit to communities large and small across the network. Information retrieval experts could contribute to this realization by extending research related to "Topic-Aware Expert Ranking" \citep{wang2013expertrank} using advanced NLP techniques that better identify knowledge on platforms where signals are likely to be quite noisy.

Additionally, these results should lead to productive conversations around when it may be reasonable to sacrifice network-wide benefit in favor of promoting equity amongst groups from AI-assisted fact-checking. Many ethical arguments could be made, and such arguments will likely rely on assumptions regarding who constitutes the communities in the network, as well as their impactedness towards certain topics. Therefore, behavioral and ethnographic research can help to demarcate which online communities are impacted by certain topics.

\subsection{Concluding Remarks}

We have shown that benefits from AI-assisted fact-checking may be unequally received amongst online communities, implying that mis- and disinformation that disproportionately impacts minority communities may indeed be falling through the cracks. Through a careful reflection on values and analysis of online communities, fact-checking organizations and platforms can build AI-assisted fact-checking systems that provide more benefit to all users.

%% file: appendix.tex
\section{Attribute and Feature Tables}
\label{sec:appendix_features}
\normalsize
Table \ref{tab:node_attr} contains the full set of node attributes and Tables \ref{tab:origin_feats} and \ref{tab:time_feats} contain the full list of features used for training check-worthiness models (described in section \ref{sec:approach_models}). Note that each observation in the check-worthiness model is a claim that has been tweeted in the simulation. Table \ref{tab:origin_feats} contains features that describe the average characteristics of all nodes that tweeted (generated utterances) about the claim, while Table \ref{tab:time_feats} contains features describing the propagation of each claim at various time points.

\begin{table}[ht]
\resizebox{\textwidth}{!}{\begin{tabular}{@{}ll@{}}
\toprule
Node Attribute & Description                                                         \\ \midrule
Impactedness & The node's impactedness values towards each topic.          \\
Belief       & The node's belief values towards each topic.                \\
Lambda              & Determines how often node is active on platform.                    \\
Wake                & Determines whether node is active on platform at current time step. \\
Inbox               & A list of all tweets that have been propagated to the node at previous time step. \\
Kind              & Identifies the node as either a normal node, a bot, or a beacon.    \\
Num\_Read     & The total number of tweets the node read for each topic.    \\ 
Prestige      & Number of followers node has    \\ \bottomrule
\end{tabular}}
\caption{TACIT simulation node attributes with descriptions}
\label{tab:node_attr}
\end{table}

\begin{table}[ht]
\center
\resizebox{\textwidth}{!}{\begin{tabular}{@{}ll@{}}
\toprule
Feature Name      & Description                                                     \\ \midrule
Avg. Origin Node Degree     & The average node in-degree (number of followers) of nodes that generated utterances about this claim. \\
Max. Origin Node Degree     & The max node in-degree (number of followers) of nodes that generated utterances about this claim.     \\
Avg. Origin Node Centrality & The average betweenness-centrality of nodes that generated utterances about this claim.               \\
Max. Origin Node Centrality & The max betweenness-centrality of nodes that generated utterances about this claim.                   \\ \bottomrule
\normalsize
\end{tabular}}
\caption{Claim Generation Features}
\label{tab:origin_feats}
\end{table}

\begin{table*}[ht]
\center
\resizebox{\textwidth}{!}{\begin{tabular}{@{}ll@{}}
\toprule
Time Step Feature Name                & Description                                                                                        \\ \midrule
Number of Nodes Visited by step t.    & The average number of reads per utterance related to this claim by time step t.                         \\
Avg. Visit Node Degree by step t.   & The average node in-degree (number of followers) of nodes that read utterances about this claim by time step t. \\
Max Visit Node Degree by step t.    & The max node in-degree (number of followers) of nodes that read utterances about this claim by time step t.     \\
Avg. Visit Node Centrality by step t. & The average betweenness-centrality of nodes that read utterances about this claim by time step t.  \\
Max Visit Node Centrality by step t.  & The max betweenness-centrality of nodes that generated utterances about this claim by time step t. \\
Max Depth from Origin by step t.      & The maximum distance from origin reached by this claim by time step t.                             \\
Nodes visited at depth d by step t. & The average number of reads per utterance about this claim at a distance d from the origin node by time step t. \\
\bottomrule
\normalsize
\end{tabular}}
\caption{Claim Propagation Features}
\label{tab:time_feats}
\end{table*}

\clearpage 

\section{Additional Results}

\begin{figure*}[ht]
 \centering
    \includegraphics[width=\textwidth]{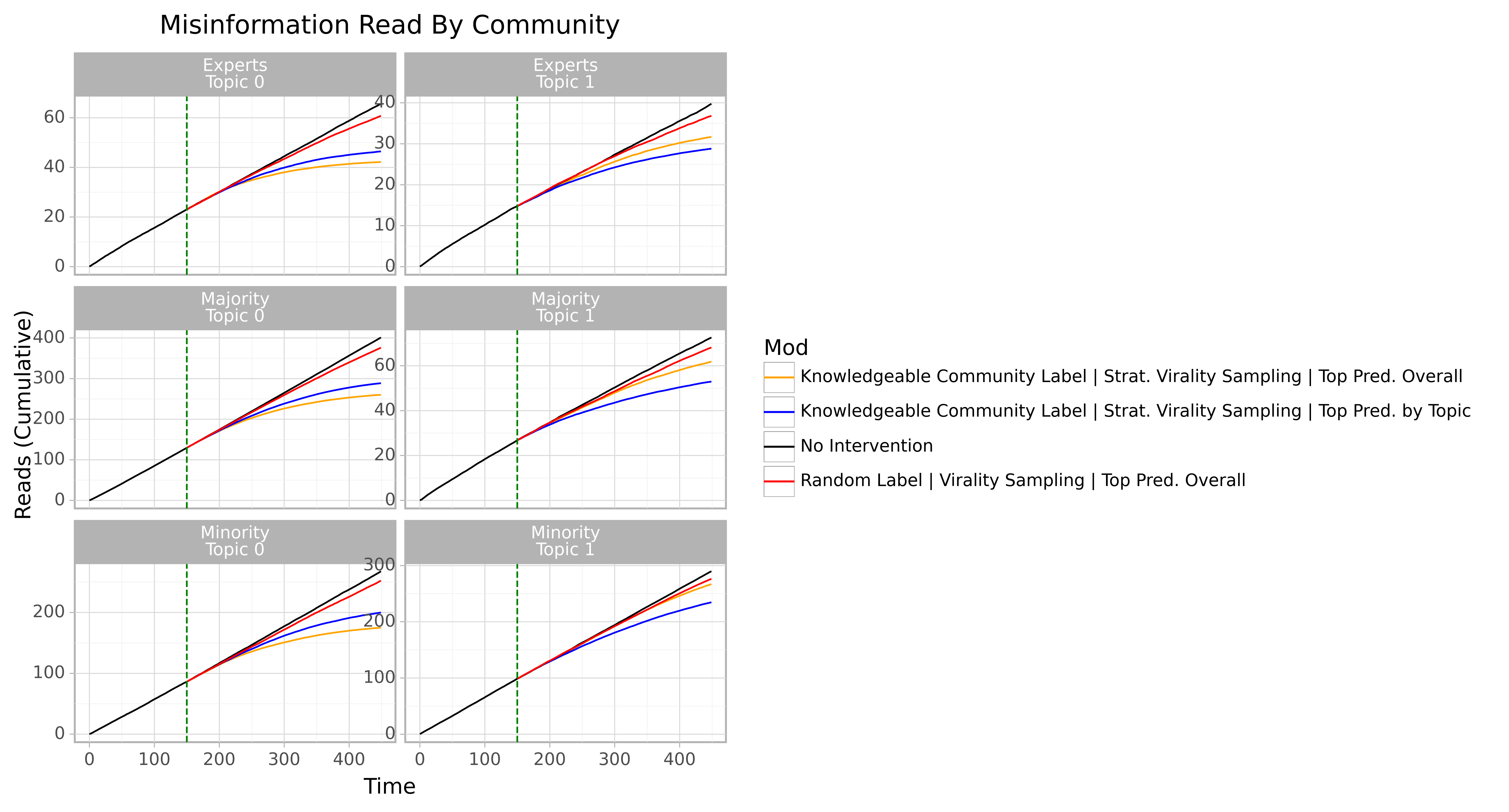}
    \caption{\textbf{Misinformation Read Over Time.} These plots show the total misinformation read by the three communities in our study (Majority, Minority, and Expert) per node over time under three different mitigations. Top Predicted By Topic yields the best reduction in misinformation read for for Topic 1 (high impact to minorities) and Top Predicted Overall yields the best reduction in misinformation read for Topic 0 (high impact to majority). This shows the effect of stratifying fact-checkers' workflow by topic: the minority community makes sizeable reductions in misinformation read about the topic they care about compared to other leading methods.}
    \label{fig:misinfo_read_over_time}
\end{figure*}

We show the amount of misinformation read by each community, over time, in Figure \ref{fig:misinfo_read_over_time} 

